\documentclass[twocolumn,10pt]{dslabarticle}

\usepackage{amsmath}
\usepackage{amsthm}
\usepackage{amssymb}

\usepackage[utf8]{inputenc}

\usepackage{dblfloatfix}

\title{From Sleep Staging to Spindle Detection: A Case Study on End-to-End Automated Sleep Analysis}
\author[1,2,3]{Niklas Grieger}
\author[2]{Siamak Mehrkanoon}
\author[4]{Philipp Ritter}
\author[1,3]{Stephan Bialonski}

\affiliation[1]{Department of Medical Engineering and Technomathematics, FH Aachen University of Applied Sciences, 52428 Jülich, Germany}
\affiliation[2]{Department of Information and Computing Sciences, Utrecht University, Utrecht, The Netherlands}
\affiliation[3]{Institute for Data-Driven Technologies, FH Aachen University of Applied Sciences, 52428 Jülich, Germany}
\affiliation[4]{Department of Psychiatry and Psychotherapy, University Hospital Carl Gustav Carus, Technische Universität Dresden, 01307 Dresden, Germany}

\metadata[Model and code]{
    \href{https://github.com/dslaborg/sumov2}{https://github.com/dslaborg/sumov2}
}
\metadata[SomnoBot]{
    \href{https://somnobot.fh-aachen.de}{https://somnobot.fh-aachen.de}
}
\metadata[Note]{
    This is a preprint of an article published in Scientific Reports 16, 16014 (2026).
    The final authenticated version is available online at: \href{https://doi.org/10.1038/s41598-026-53891-9}{https://doi.org/10.1038/s41598-026-53891-9}
}

\abstract{
    Automation of sleep analysis, including both macrostructural (sleep stages) and microstructural (e.g., sleep spindles) elements, promises to enable large-scale sleep studies and to reduce variance due to inter-rater incongruencies.
    While individual steps, such as sleep staging and spindle detection, have been studied separately, the feasibility of automating multi-step sleep analysis remains unclear.
    In this case study, we evaluate whether a fully automated analysis using validated machine learning models for sleep staging (RobustSleepNet) and subsequent spindle detection (SUMOv2) can replicate findings from an expert-based study of bipolar disorder.
    The automated analysis qualitatively reproduced key findings from the expert-based study, including significant differences in fast spindle densities between bipolar patients and healthy controls, accomplishing in minutes what previously took months to complete manually.
    While the results of the automated analysis differed quantitatively from the expert-based study, possibly due to biases between expert raters or between raters and the models, the models individually performed at or above inter-rater agreement for both sleep staging and spindle detection.
    Our results demonstrate that fully automated approaches have the potential to facilitate large-scale sleep research.
    We are providing public access to the tools used in our automated analysis by sharing our code and introducing SomnoBot, a privacy-preserving sleep analysis platform.
}

\begin{document}
\maketitle

\section{Introduction}

Automating EEG sleep analysis has the potential to enable scalable and cost-effective studies that were previously impractical, thereby creating opportunities to gain new insights into a wide range of diseases that interact with sleep (such as affective disorders).
Recent decades have seen substantial progress in automating various aspects of sleep analysis.
These advancements range from the characterization of sleep macroarchitecture, involving the detection of sleep stages~\cite{Gaiduk2023,Phan2022,Fiorillo2019}, to the identification of EEG graphoelements (sleep microarchitecture)~\cite{Hermans2022} such as sleep spindles~\cite{CoppieterstWallant2016}, K-complexes~\cite{Tapia2020}, or rapid-eye movements~\cite{Yetton2016}, which reflect intricate processes at faster temporal scales occurring in the brain during sleep.
While early approaches for automated analyses relied upon defining features such as power in spectral bands to indicate particular sleep stages or sleep spindles, recent methods primarily leverage deep neural network models capable of automatically learning distinctive features to achieve state-of-the-art detection performances~\cite{Phan2022}.

\begin{figure*}
    \centering
    \includegraphics[width=\linewidth]{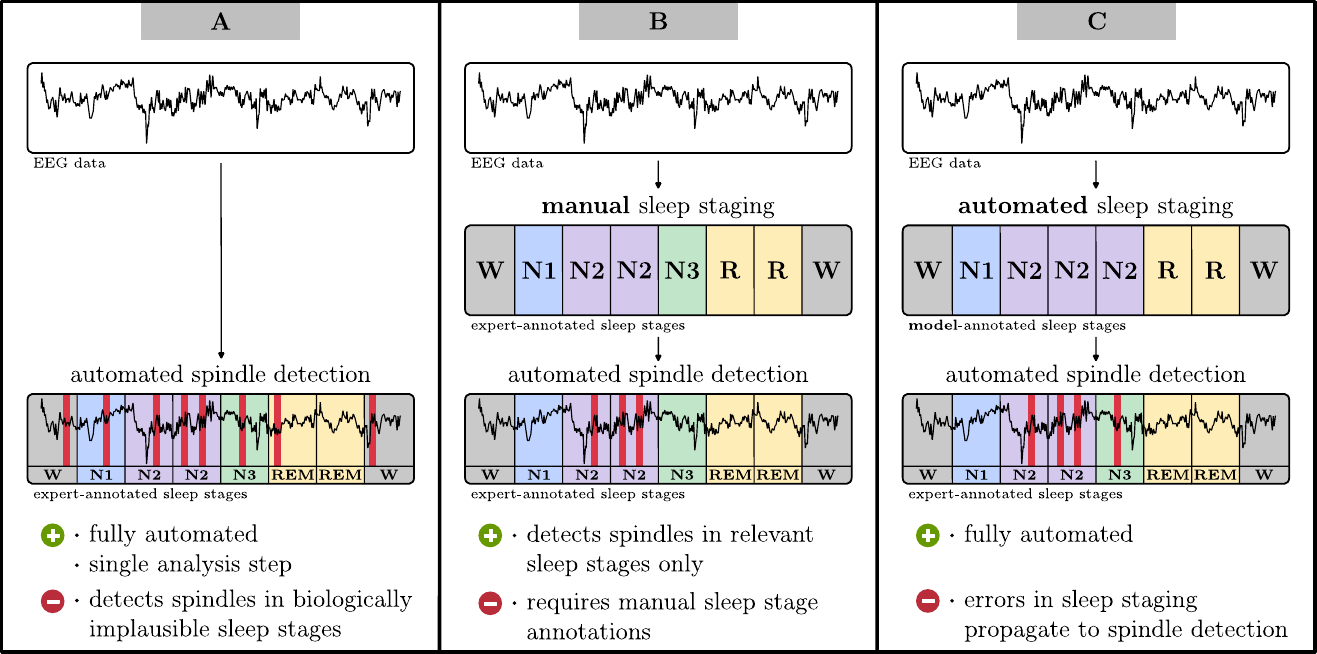}
    \caption{
        Overview of usual sleep analysis approaches with automated spindle detection in N2 sleep.
        Applying a spindle detection model to EEG data without accounting for sleep stages (A) is a simple and fully automated approach but may lead to spindles being detected in Wake or REM sleep stages, where they are biologically implausible.
        More sophisticated approaches involve first annotating sleep stages, either manually (B) or using an automated sleep staging model (C), to restrict spindle detection to N2 (or, additionally, N1 and N3) sleep stages.
    }
    \label{fig:pipeline}
\end{figure*}

Assessing the performance of models is challenging since their sleep annotations need to be compared with expert annotations that are known to vary between experts (inter-rater variability) and within the same expert over time (intra-rater variability).
Inter-rater variability can arise from subjective biases or differing practices in labs and clinics, despite experts' adherence to annotation guidelines, such as the American Academy of Sleep Medicine (AASM) scoring rules~\cite{Berry2020} or the Rechtschaffen and Kales (R\&K) criteria~\cite{Rechtschaffen1968}.
Given this variability, one strategy for assessing a model's performance is to compare it against consensus annotations created by a group of experts.
Such a consensus can provide a more reliable reference for evaluating model performance, and training models to mimic a consensus has been shown to yield annotations that are comparable to or even better than the agreement between individual expert scorers and the consensus~\cite{Bakker2022}.
Another approach for evaluating model performance is based on comparing the agreement between a model and an expert scorer (model-expert agreement) to a distribution of agreements between pairs of experts (inter-rater agreement distribution).
If the model-expert agreement is similar to or better than the average inter-rater agreement between pairs of expert scorers, the model's performance can be considered robust and comparable to that of an expert.

Given these strategies to assess model performance, it has been shown that individual steps of sleep analysis like sleep staging and spindle detection can be automated with high accuracy~\cite{Guillot2021,Perslev2021,Olesen2020,Vallat2021,Hanna2023,Kaulen2022}.
However, these steps have been evaluated separately, and it remains unclear whether a multi-step sleep analysis process can be automated end-to-end while still delivering reliable sleep metrics precise enough to test scientific hypotheses or even support clinical diagnostics.
In such end-to-end analyses, various interdependencies between the individual steps must be taken into consideration.
For instance, the automated detection of sleep spindles is commonly constrained to specific sleep stages (see Fig.~\ref{fig:pipeline}), such as N2 (and occasionally N3) non-REM sleep, and thus relies on the accurate detection of sleep stages.
This limitation arises because models for sleep spindle detection are usually trained on datasets (training sets) containing only those stages where spindles are biologically plausible under AASM scoring rules.
When presented with stages outside their training set, such as REM sleep or wakefulness (Wake), that are present in usual overnight recordings, models will often detect spindles in sleep stages where such events should not occur (see Fig.~\ref{fig:pipeline}A).
This challenge extends across methods for analyzing sleep microstructure, including K-complexes, spindles, and rapid eye movements (REMS), highlighting the need for evaluations of multi-step sleep analyses end-to-end.

\begin{figure*}
    \centering
    \includegraphics[width=\textwidth]{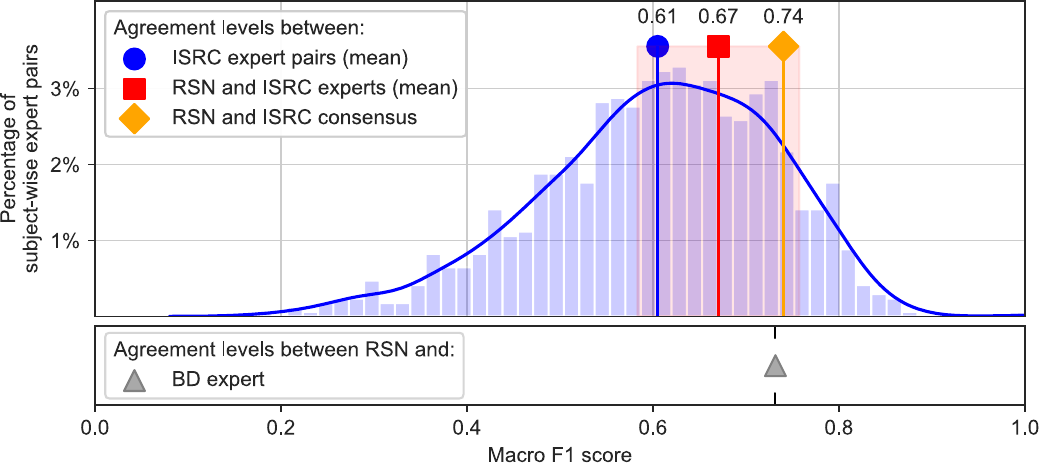}
    \caption{
        Comparison of model-expert agreement for sleep staging with inter-rater agreement levels among human experts on the IS-RC dataset.
        The average agreement between the RSN model and the IS-RC experts (quantified by mean macro F1 score, x-axis) is shown as a red square ($\blacksquare$), with the shaded region denoting the standard deviation across experts and recordings.
        The diamond symbol ($\blacklozenge$) indicates the agreement between the RSN model and a consensus of all IS-RC experts.
        To compare the model-expert agreement with the agreement between pairs of expert scorers, the histogram (blue bars) and its density estimate (blue curve) show the distribution of macro F1 scores calculated for all subject-wise comparisons of expert pairs in the IS-RC dataset.
        The blue circle ({\Large$\bullet$}) shows the average of this distribution.
        For context, we also show the average model-expert agreement between RSN and the BD expert ($\blacktriangle$).
    }
    \label{fig:sleep-stages}
\end{figure*}

In this case study, we investigated whether an automated end-to-end sleep analysis using validated deep learning models for sleep staging and subsequent spindle detection can replicate findings from a previous study in which data was manually annotated by expert scorers.
This prior study compared spindle metrics between individuals with bipolar disorder (BD) and healthy controls and contained data that the sleep analysis models have not encountered before, emulating a real-world scenario.
Beyond replicating the expert-led findings, we validated each step of the automated analysis on separate datasets annotated by multiple experts to compare model-expert agreements with distributions of inter-rater agreements.
Our findings indicate that automated multi-step analyses can qualitatively reproduce key differences and similarities in spindle characteristics between bipolar patients and healthy controls, consistent with the original expert-led study.
These results highlight the potential of fully automated analyses to replicate expert-led analyses, paving the way for scalable, reproducible sleep research.

We have made the code used for the automated sleep analysis, including our enhanced spindle detection model (SUMOv2), publicly available online.
Additionally, we are providing public access to SomnoBot (\url{https://somnobot.fh-aachen.de}), a privacy-preserving tool that allows researchers to automatically analyze their data without requiring programming skills or sharing data with third parties.

\section{Results}
\label{sec:results}

Our automated end-to-end sleep analysis used two deep neural network models for sleep staging and spindle detection, respectively (see Fig.~\ref{fig:pipeline}C).
We evaluated each model by comparing its agreement with expert annotations to the distribution of inter-rater agreements, which we derived from EEG datasets annotated by multiple expert scorers.
Finally, we leveraged the automated analysis process to retrospectively analyze a dataset on bipolar disorder (BD) and examine whether it could replicate findings from a previous expert-led study~\cite{Ritter2018}, thus removing the need for manual scoring.

In that study, a human expert manually annotated sleep stages and sleep spindles following AASM guidelines in overnight polysomnography (PSG) recordings from 25 healthy controls and 23 bipolar disorder patients~\cite{Ritter2018}.
The study's main findings included a significantly lower density of fast sleep spindles in bipolar subjects compared to healthy controls, suggesting that this feature may serve as a potential biomarker -- an intriguing finding given the current lack of reliable biomarkers for bipolar disorder.

\subsection{Sleep Staging Performance}

\begin{figure*}
    \centering
    \includegraphics[width=\textwidth]{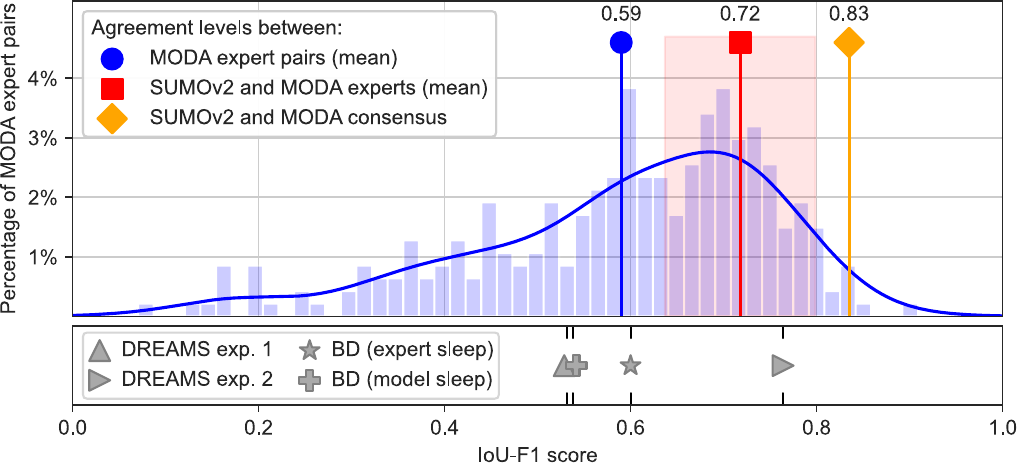}
    \caption{
        Comparison of model-expert agreement for spindle detection with inter-rater agreement among human annotators on the MODA dataset.
        The red square ($\blacksquare$) and orange diamond ($\blacklozenge$) symbols indicate the agreement between the SUMOv2 model and individual MODA experts (STD as shaded area) or a consensus of all MODA experts, respectively.
        Inter-agreement levels between expert pairs on the MODA dataset are shown in the histogram (blue bars) and its density estimate (blue curve).
        The average of this distribution is shown as a blue circle ({\Large$\bullet$}).
        To contextualize these results across different datasets and experts, we show the average model-expert agreement between SUMOv2 and the two DREAMS experts ($\blacktriangle$ and \rotatebox[origin=c]{270}{$\blacktriangle$}), as well as between SUMOv2 and the BD expert when using expert-annotated N2 sleep stages ($\star$) or N2 sleep stages detected by the RSN model ($+$), respectively.
    }
    \label{fig:spindles}
\end{figure*}

We automatically detected sleep stages in the BD study recordings using the RobustSleepNet (RSN) model, which has previously demonstrated strong sleep staging performance~\cite{Guillot2021}.
Notably, the RSN model has never been trained or evaluated on BD recordings, reflecting real-world scenarios, where machine learning models must analyze previously unseen patient data.
To compare the RSN model to an expert scorer, we calculated the macro F1 (MF1) score, which ranges from 0 (indicating no agreement) to 1 (perfect agreement).
MF1 agreement scores were calculated for each of the 48 BD recordings, obtaining values ranging from 0.68 (first quartile) to 0.79 (third quartile).
We observed the average of the MF1 scores to be $0.73$ with a standard deviation (SD) of $0.11$.

To contextualize the obtained MF1 scores, we quantified inter-rater agreement on the IS-RC dataset, a publicly available dataset consisting of 69 overnight recordings that were scored independently by six experts~\cite{Kuna2013}.
For each pair of experts and each recording, we determined an MF1 score to measure the inter-rater agreement level and show the distribution of these scores in Fig.~\ref{fig:sleep-stages} (blue bars).
Pairs of human expert scorers reached an average MF1 score of $0.61$ (SD = $0.10$), with MF1 values ranging from $0.54$ (first quartile) to $0.68$ (third quartile).
Additionally, we calculated MF1 scores between the RSN model and the individual IS-RC experts, obtaining an average of $0.67$ (SD = $0.09$), significantly higher than the MF1 scores observed between expert pairs ($p<0.001$, one-sided t-test with independent samples).
In line with previous findings~\cite{Stephansen2018}, model performance increased when evaluating RSN against a consensus of multiple experts rather than against single scorers.

\subsection{Sleep Spindle Detection Performance}

We detected sleep spindles with the SUMOv2 model (see Methods), which improves upon the publicly available machine learning model SUMO that achieved state-of-the-art detection performance in a previous study~\cite{Kaulen2022}.
Developed without prior exposure to the BD recordings, SUMOv2 was evaluated for its agreement with an expert scorer by calculating the Intersection-over-Union F1 (IoU-F1) score, requiring at least 20\% overlap between an expert- and model-identified spindle to be considered a valid agreement.
The IoU-F1 score obtains values between 0 (indicating no agreement) and 1 (perfect agreement).
To assess SUMOv2 independently of automated sleep staging, we evaluated its spindle detection performance in N2 sleep epochs identified by the expert scorer (see Fig.~\ref{fig:pipeline}B), where it achieved an IoU-F1 score of $0.60$ in the BD recordings.
When sleep stages were automatically detected by the RSN model, thereby enabling a fully automated analysis process (Fig.~\ref{fig:pipeline}C), the agreement between SUMOv2 and expert-detected spindles decreased slightly, yielding an IoU-F1 of $0.54$.

To gain an understanding of the reliability of SUMOv2's performance, we evaluated additional datasets annotated by different expert scorers: DREAMS (sleep spindles from 8 subjects, annotated by two experts~\cite{Devuyst2005}) and MODA (artifact-free N2 sleep spindles from 180 subjects, annotated by 47 experts~\cite{Lacourse2020}).
On DREAMS, SUMOv2 achieved IoU-F1 scores of $0.53$ and $0.76$ when evaluated against the first and second expert, respectively, demonstrating that model-expert agreement can vary substantially across individual raters.

The larger number of expert scorers in MODA allowed us to compare SUMOv2 to inter-rater agreement levels observed between expert pairs.
On the MODA subset not used for developing SUMOv2, the model achieved an average IoU-F1 score of $0.72$ (SD = $0.08$, shaded area in Fig.~\ref{fig:spindles}) when evaluated against individual experts.
On the same data, the average inter-rater agreement across expert pairs was lower, with a mean IoU-F1 of $0.59$ (SD = $0.16$), and expert pair scores ranging from $0.50$ (first quartile) to $0.71$ (third quartile; see histogram in Fig.~\ref{fig:spindles}).
Furthermore, when evaluated against MODA's consensus annotations---where spindles were only valid if identified independently by multiple experts---SUMOv2 achieved an IoU-F1 of $0.83$.

\subsection{Spindle Characteristics of Bipolar and Healthy Subjects}

We performed a fully automated analysis of the BD recordings by first identifying sleep stages using the RSN model, followed by detecting sleep spindles in N2 sleep epochs from frontal and central channels with the SUMOv2 model (see Fig.~\ref{fig:pipeline}C).
Following the original analysis in the BD study~\cite{Ritter2018}, we calculated for each detected spindle its duration, dominant frequency, and amplitude.
Then we determined the average spindle characteristics and spindle density separately for fast spindles (dominant frequency $>13$~Hz) and slow spindles ($\leq$ 13~Hz) in both the healthy and bipolar cohort (see Tables~\ref{tab:spindle-features-fast} and \ref{tab:spindle-features-slow}).

\begin{figure}
    \centering
    \includegraphics[width=\linewidth]{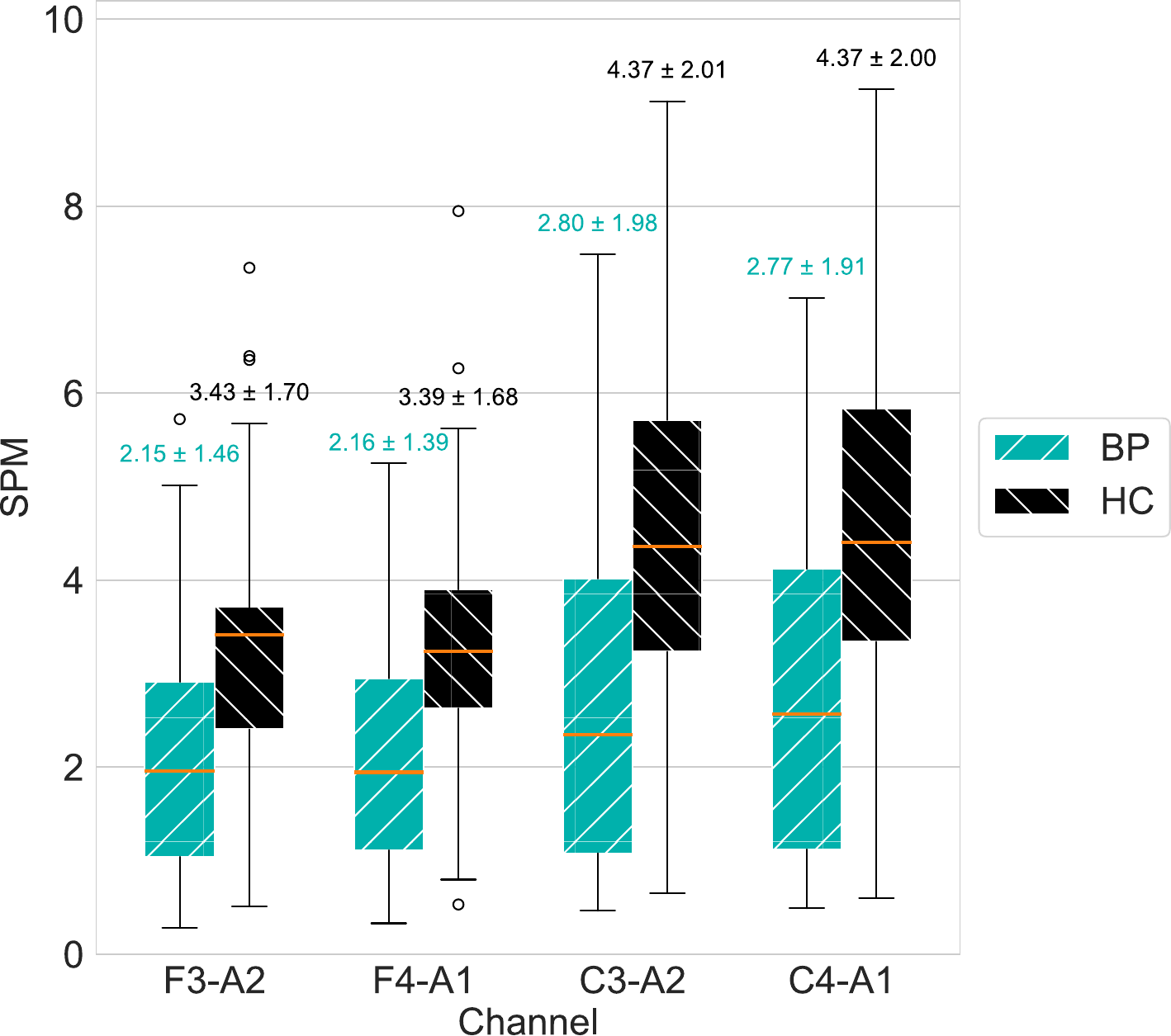}
    \caption{
        Boxplot of spindle densities for fast spindles (frequencies $>$ 13~Hz) as detected by the SUMOv2 model in the BD recordings for healthy controls (HC) and patients with bipolar disorder (BP).
        Spindles were detected in N2 sleep annotated by the RSN model.
        The labels above the boxplots indicate the mean and standard deviation of the spindles per minute (SPM) for each cohort and channel.
    }
    \label{fig:nikitin-spms}
\end{figure}

We observed significant differences in spindle densities of fast spindles between healthy controls and bipolar subjects (see Fig.~\ref{fig:nikitin-spms} and Table~\ref{tab:spindle-features-fast}).
These differences were consistent across frontal and central EEG channels, with average spindle densities of 2.16 spindles per minute (SPM) for bipolar patients (BP) and 3.41 SPM for healthy controls (HC) in frontal channels, and 2.79 SPM (BP) and 4.37 SPM (HC) in central channels.
Spindle densities varied substantially between individual subjects, both for healthy controls (standard deviations between 1.68 and 2.01) and bipolar patients (standard deviations between 1.39 and 1.98).
We found no statistically significant differences in slow spindle densities or other characteristics between the two cohorts (see Tables~\ref{tab:spindle-features-fast} and~\ref{tab:spindle-features-slow}).

When analyzing spindle characteristics in regard to channel placement, fast spindles showed higher densities and amplitudes in central channels compared to frontal channels, both in bipolar subjects (density: 2.79 vs. 2.16 SPM; amplitude: 8.40 vs. 7.03~$\mu$V) and healthy controls (density: 4.37 vs. 3.41 SPM; amplitude: 8.39 vs. 7.04~$\mu$V).
In contrast, slow spindles demonstrated slightly higher densities and amplitudes in frontal channels for bipolar patients (density: 3.38 SPM vs. 2.75 SPM; amplitude: 9.14~$\mu$V vs. 8.79~$\mu$V) and healthy controls (density: 3.76 SPM vs. 2.80 SPM; amplitude: 9.37~$\mu$V vs. 8.63~$\mu$V).
Spindle durations remained stable across channels, averaging 0.82--0.88~s, while average frequencies were consistent for both fast (13.88--13.98~Hz) and slow spindles (12.16--12.27~Hz).

\begin{table}[h]
    \centering
    \footnotesize
    \setlength{\tabcolsep}{2.5pt}
    \begin{tabular}{@{}lllll@{}}
                  & Channel & HC           & BP           & p-value      \\
        \hline
        \hline
        Density   & F3-A2   & 3.43 (1.70)  & 2.15 (1.46)  & 0.008        \\
                  & F4-A1   & 3.39 (1.68)  & 2.16 (1.39)  & 0.009        \\
                  & C3-A2   & 4.37 (2.01)  & 2.80 (1.98)  & 0.011        \\
                  & C4-A1   & 4.37 (2.00)  & 2.77 (1.91)  & 0.008        \\
        \hline
        Frequency & F3-A2   & 13.91 (0.14) & 13.91 (0.18) & 0.990        \\
                  & F4-A1   & 13.92 (0.13) & 13.92 (0.12) & 0.983        \\
                  & C3-A2   & 13.98 (0.22) & 13.90 (0.18) & 0.184        \\
                  & C4-A1   & 13.98 (0.21) & 13.88 (0.17) & 0.089        \\
        \hline
        Duration  & F3-A2   & 0.87 (0.09)  & 0.82 (0.12)  & 0.125        \\
                  & F4-A1   & 0.87 (0.09)  & 0.82 (0.12)  & 0.150        \\
                  & C3-A2   & 0.88 (0.09)  & 0.83 (0.13)  & 0.154        \\
                  & C4-A1   & 0.88 (0.10)  & 0.83 (0.13)  & 0.203        \\
        \hline
        Amplitude & F3-A2   & 6.95 (1.49)  & 6.99 (1.43)  & 0.910$^\dag$ \\
                  & F4-A1   & 7.12 (1.40)  & 7.06 (1.39)  & 0.992$^\dag$ \\
                  & C3-A2   & 8.36 (1.73)  & 8.32 (2.13)  & 0.703$^\dag$ \\
                  & C4-A1   & 8.41 (1.88)  & 8.47 (1.61)  & 0.900        \\
        \hline
    \end{tabular}
    \caption{
        Characteristics of fast spindles (frequency $>$ 13~Hz) detected by the SUMOv2 model in the BD recordings for healthy controls (HC) and patients with bipolar disorder (BP).
        While spindles were aggregated across all channels, we calculated individual spindle characteristics separately for each channel using only that channel's EEG signal.
        P-values were calculated between the characteristics of the two cohorts with Welch's two-sided independent t-tests unless indicated otherwise.
        We use $\dag$ to indicate that a Wilcoxon two-sided rank-sum test was used.
    }
    \label{tab:spindle-features-fast}
\end{table}

\begin{table}[h]
    \centering
    \footnotesize
    \setlength{\tabcolsep}{2.5pt}
    \begin{tabular}{@{}lllll@{}}
                  & Channel & HC           & BP           & p-value      \\
        \hline
        \hline
        Density   & F3-A2   & 3.74 (1.87)  & 3.38 (2.15)  & 0.547        \\
                  & F4-A1   & 3.78 (1.85)  & 3.37 (2.12)  & 0.337$^\dag$ \\
                  & C3-A2   & 2.80 (1.57)  & 2.73 (1.65)  & 0.870        \\
                  & C4-A1   & 2.80 (1.57)  & 2.76 (1.61)  & 0.925        \\
        \hline
        Frequency & F3-A2   & 12.26 (0.15) & 12.16 (0.23) & 0.116        \\
                  & F4-A1   & 12.25 (0.14) & 12.17 (0.22) & 0.159        \\
                  & C3-A2   & 12.27 (0.11) & 12.19 (0.20) & 0.111        \\
                  & C4-A1   & 12.26 (0.11) & 12.18 (0.20) & 0.109        \\
        \hline
        Duration  & F3-A2   & 0.88 (0.11)  & 0.85 (0.13)  & 0.392        \\
                  & F4-A1   & 0.88 (0.11)  & 0.84 (0.13)  & 0.350        \\
                  & C3-A2   & 0.86 (0.11)  & 0.84 (0.12)  & 0.468        \\
                  & C4-A1   & 0.87 (0.11)  & 0.84 (0.13)  & 0.403        \\
        \hline
        Amplitude & F3-A2   & 9.29 (1.51)  & 9.21 (1.74)  & 0.599$^\dag$ \\
                  & F4-A1   & 9.45 (1.48)  & 9.06 (1.57)  & 0.397        \\
                  & C3-A2   & 8.60 (1.80)  & 8.67 (1.74)  & 0.897        \\
                  & C4-A1   & 8.66 (1.79)  & 8.90 (1.50)  & 0.635        \\
        \hline
    \end{tabular}
    \caption{
        Characteristics of slow spindles (frequency $\le$ 13~Hz) detected by the SUMOv2 model in the BD recordings for healthy controls (HC) and patients with bipolar disorder (BP).
        While spindles were aggregated across all channels, we calculated individual spindle characteristics separately for each channel using only that channel's EEG signal.
        P-values were calculated between the characteristics of the two cohorts with Welch's two-sided independent t-tests unless indicated otherwise.
        We use $\dag$ to indicate that a Wilcoxon two-sided rank-sum test was used.
    }
    \label{tab:spindle-features-slow}
\end{table}

\section{Discussion}

We observed the automated analysis via deep learning models to qualitatively replicate key findings of a study on bipolar disorder by Ritter et al.~\cite{Ritter2018}, achieving in minutes what had previously required months to complete manually.
Our approach found significant differences in fast spindle densities between healthy controls and patients with bipolar disorder, consistent with the prior study~\cite{Ritter2018} and similar to observations made for schizophrenia~\cite{Ferrarelli2021} (see Fig.~\ref{fig:nikitin-spms}).
Our analysis found systematically higher spindle densities than those observed in the expert annotations~\cite{Ritter2018} (5.53$\pm$3.20 vs. 3.49$\pm$2.04 SPM for patients, 7.17$\pm$2.78 vs 4.23$\pm$1.79 SPM for controls).
This discrepancy might reflect a bias between the model and the expert, akin to those found among individual expert scorers (see the broad distribution of inter-rater agreement in Fig.~\ref{fig:spindles}).
However, we consider it more likely that the difference is explained by differences in the spindle aggregation methods used in our analysis of the BD dataset (see section~\ref{ssec:methods-spindles}), a conclusion further supported by our model's predictions on other single-channel EEG datasets, which yielded lower spindle densities (DREAMS: 2.50 $\pm$ 1.28 SPM, MODA: 3.88 $\pm$ 3.02 SPM).

Our automated analysis was also able to replicate the second finding of the original study that average spindle frequencies for both fast and slow spindles were slightly lower in bipolar subjects than in healthy controls.
While the prior study found significant difference for frequencies of fast spindles in central channels ($p<0.02$), our analysis showed a similar trend but without statistical significance ($p<0.19$, see Tab.~\ref{tab:spindle-features-fast}), with the smallest p-values also observed in central channels.

We found differences in the absolute values of spindle durations, frequencies and amplitudes compared to the original study (see Tables~\ref{tab:spindle-features-fast} and~\ref{tab:spindle-features-slow}).
Our detected spindles were generally shorter (less than 0.9~s compared to more than 1~s in the original study), and showed slight variations in frequency (13.9--14.0~Hz in our results vs. 13.5--13.7~Hz for fast spindles, with comparable slow spindle frequencies) and amplitude (7.0--8.5~$\mu$V in our results  vs. 8.3--10.4~$\mu$V for fast spindles, 8.6--9.5~$\mu$V vs. 9.5--10.3~$\mu$V for slow spindles).
These discrepancies can likely be attributed to three main factors: differences in calculation methods for amplitude and frequency, known variability in spindle duration assessments between experts~\cite{Wendt2015}, and potential dataset bias, as our SUMOv2 model was trained on the MODA dataset where spindles tend to be shorter (0.75--0.79~s~\cite{Lacourse2020}) compared to those in the BD dataset.

Analyzing the model-expert agreement at each of the two analysis steps, we found strong agreement between automatically detected and expert-annotated sleep stages, consistent with previous studies across various datasets~\cite{Guillot2021,Perslev2021,Olesen2020,Vallat2021,Hanna2023}.
The RSN model achieved a high macro F1 score of 0.73 on the BD dataset, which aligns with the range of inter-rater agreements observed between expert pairs on the IS-RC dataset (see Fig.~\ref{fig:sleep-stages}).
When we compared this inter-rater agreement with the agreement between RSN and the IS-RC experts on the same data, RSN reached significantly higher macro F1 scores than the expert pairs ($p < 0.001$).
We further found modest variation in the agreements between expert pairs (first and third quartiles: 0.54--0.68), in line with previously reported inter-rater agreements of 0.54--0.86 (Cohen's Kappa)~\cite{DankerHopfe2004,DankerHopfe2009,Silber2007,Lee2022,Zhang2014,Basner2008}.

For spindle detection, the SUMOv2 model achieved high IoU-F1 scores on the BD and DREAMS data, indicating that it can reliably identify spindles in datasets outside the training data, a capability that has rarely been studied in previous work~\cite{Chambon2019,Kulkarni2019,You2021}.
When sleep stages were provided by an expert (see Fig.~\ref{fig:pipeline}B), SUMOv2 achieved IoU-F1 scores of 0.60 for the BD and 0.53 and 0.76 for the DREAMS dataset.
When sleep stages were instead automatically detected by RSN (see Fig.~\ref{fig:pipeline}C), SUMOv2's agreement with the BD expert decreased slightly to an IoU-F1 of 0.54.
In both cases, the model-expert agreements lay within the range of expert pair agreements observed on the MODA dataset (see Fig.~\ref{fig:spindles}) and within the typical agreements of 0.42--0.61 (IoU-F1 scores) reported in the literature~\cite{Wendt2015,Tamamoto2024,Devuyst2005}.
Although this comparison should be interpreted with caution due to differences in datasets and annotators, it provides a sense of how SUMOv2 performs in real-world scenarios with new data.
In order to establish an unbiased comparison between SUMOv2 and the MODA expert pairs on a shared set of data and scorers, we also evaluated SUMOv2 against individual experts on a MODA test dataset not used for model training.
On this dataset, SUMOv2 achieved substantially higher IoU-F1 scores than those observed between the expert pairs.
While this comparison only partially reflects real-world scenarios due to the same experts annotating both training and test data, it is nevertheless an indicator of SUMOv2's performance compared to typical inter-rater agreements one could expect in practice.

Despite the promising performance of our fully automated sleep analysis approach, our study has several limitations.
First, publicly accessible datasets with jointly annotated sleep stages and spindles are scarce, limiting our study to those available.
For this reason, we were able to reproduce only a single previously published study, rather than validating our approach across multiple datasets, which would have provided stronger evidence for generalizability.
While we were able to include additional datasets in evaluating the individual analysis steps, these datasets were limited in size and diversity, and may not fully represent the broader population.
This limitation could impact our comparisons of model-expert agreement with inter-rater agreement scores, as sleep macro- and microarchitecture can vary substantially with factors such as age, sex, and various physiological or pathological conditions~\cite{Kocevska2020}, all of which may influence the achievable level of agreement among experts.
Future research could benefit from expanding investigations to more diverse populations, including patients with specific disorders or recordings from mobile EEG devices, and exploring performance stratification by demographic factors.
Achieving these goals will require the availability of large, well-annotated datasets and the establishment of a common benchmark to support the development and evaluation of fully automated sleep analysis approaches within the research community.
Second, our analysis approach did not incorporate explicit handling of EEG artifacts, which we suggest as a consideration for future research.
Since artifacts can distort EEG recordings, integrating an automated artifact detection mechanism could enhance the robustness of our sleep analysis approach.
We note, however, that most state-of-the-art sleep analysis models are trained on EEG data that already includes artifacts, potentially enabling them to adapt and mitigate their effects implicitly.
Third, our study focused on a sequential analysis approach in which sleep staging precedes spindle detection rather than an end-to-end modeling approach that jointly optimizes both tasks.
While such an integrated approach could offer advantages, it requires extensive datasets with jointly annotated sleep stages and spindles, which are currently not publicly available as noted above.

Our study advances automated sleep analysis by demonstrating that fully automated sleep staging and spindle detection can match expert-level performance.
To enable broader access to automated sleep analysis, we are sharing our code, publishing our novel spindle detection model, SUMOv2, and releasing our privacy-preserving sleep analysis tool, SomnoBot (\url{https://somnobot.fh-aachen.de}), which enables researchers to use our analysis approach without requiring programming expertise.
We hope that this work and similar efforts will facilitate large-scale, long-term sleep studies, enabling new insights into sleep-related health and disease.

\section{Methods}
\label{sec:methods}

\subsection{Automatic Detection of Sleep Stages}
\label{ssec:methods-sleep-staging}

\paragraph{Datasets.}

The IS-RC dataset contains one PSG recording each of 70 women, recorded in a research study investigating sleep disordered breathing in women in midlife (40--57 years old)~\cite{Kuna2013}.
All 70 recordings were annotated by ten expert scorers following the guidelines of the American Academy of Sleep Medicine (AASM)~\cite{Berry2020}, and the annotations of six experts are publicly available.
We aggregated these annotations into a consensus following the approach outlined by Stephansen~et~al.~\cite{Stephansen2018}.
One recording was discarded due to a mismatch between the filenames of the annotations and the corresponding EEG recording.

The BD (Bipolar EEG Dataset) dataset contains one PSG recording each of 25 healthy controls and 23 patients with bipolar disorder~\cite{Ritter2018}.
The 48 recordings were annotated by a single expert following the AASM guidelines.
Due to differences in channel setup, we focused our analyses on the set of EEG channels that were common to all recordings: F3-A2, F4-A1, C3-A2, C4-A1, O1-A2, O2-A1, and A1-A2.
Sampling rates varied between 100~Hz, 200~Hz, or 500~Hz depending on the recording and channel.

All recordings were divided into 30-second epochs and labeled as either Wake, N1, N2, N3, or REM sleep stages by the expert annotators.

\paragraph{Model.}
We used the RobustSleepNet (RSN) model for automated sleep staging which is a deep learning model that was designed to be invariant to the number, type, or order of PSG montages~\cite{Guillot2021}.
Guillot~et~al. provide several checkpoints for RSN that were trained on EEG, EOG, and EMG data from different datasets~\cite{Guillot2021b}.
We used the checkpoint trained on 659 recordings from the MESA~\cite{Chen2015}, MrOS~\cite{Blackwell2011}, SHHS~\cite{Quan1997}, DODO~\cite{Guillot2020}, DODH~\cite{Guillot2020}, and MASS~\cite{OReilly2014} datasets.
The model accepts input sequences of 21 sleep epochs (i.e., 10.5 minutes) to ensure sufficient context for the classification of each epoch.
Given an input sequence, the model outputs probabilities for each of the five sleep stages for each epoch.
Due to the model's architecture, the input sequences could have an arbitrary number of channels.

Following standard procedures for the use of RSN~\cite{Guillot2021}, each recording was preprocessed before being scored by the model using a 4th order Butterworth bandpass filter (0.3--30~Hz), downsampled to 60~Hz, and normalized by subtracting the median and dividing by the interquartile range.
Amplitudes outside -20 to 20 were clipped to these bounds.
We buffered the beginning and the end of each preprocessed recording with 20 sleep epochs of zeros to prevent the model from making predictions based on incomplete sequences.
We then created the input sequences to the model by sliding a window of 21 epochs over the buffered recording with a step size of 1 epoch (i.e., each epoch was part of 21 input sequences).
The resulting 21 predicted probabilities for each sleep epoch were aggregated by calculating the geometric mean to obtain the final predictions for each epoch~\cite{Guillot2021}.

\paragraph{Evaluation.}
Given two sets of annotations for a recording, we determined the agreement between the two sets by calculating the Macro F1 score as follows.
For each sleep stage, we first counted the number of epochs that matched in both annotations as true positives (TP).
False positives (FP) were defined as epochs labeled as a given sleep stage in one annotation but assigned a different stage in the other annotation.
Conversely, false negatives (FN) were epochs that were assigned a different stage in one annotation but labeled as the given stage in the other annotation.
Precision and recall for a stage was calculated as TP / (TP + FP) and TP / (TP + FN), respectively, and the F1 score for that stage was then given by 2 $\times$ (precision $\times$ recall) / (precision + recall).
Finally, the Macro F1 score was calculated by averaging the stage-specific F1 scores.

\subsection{Automatic Detection of Sleep Spindles}
\label{ssec:methods-spindles}

We detected sleep spindles using SUMOv2, an enhanced version of the publicly available SUMO model~\cite{Kaulen2022}.
Our improvements focused on increasing robustness to variations in amplitude scales, ensuring more reliable spindle detection across diverse datasets.

\paragraph{Datasets.}
The MASS dataset contains 200 PSG recordings from 200 mostly healthy subjects (15 subjects were diagnosed with mild cognitive impairment) and sampled at 256~Hz~\cite{OReilly2014}.
The MODA dataset provides spindle annotations for selected sections of 180 recordings from MASS~\cite{Lacourse2020,Yetton2022}.
Each of these recording was divided into 10 (30 recordings) or 3 (150 recordings) blocks of 115~s of artifact-free N2 sleep.
The blocks were annotated for spindles by up to seven human experts, who used either the C3-A2 or C3-LE EEG channel for annotations, depending on the recording.
In total, the MODA dataset contains 749 blocks annotated by 47 experts (one block was not presented to any experts).
Annotations were aggregated into a consensus based on the experts' confidence in their annotations~\cite{Lacourse2020}.
For simplicity, we refer to the combination of the MASS and MODA datasets as the MODA dataset.

The DREAMS dataset comprises eight 30 minutes long segments of EEG data from eight subjects with various pathologies (dysomnia, restless legs syndrome, insomnia, apnoea/hypopnoea syndrome) that was sampled at 50, 100, or 200~Hz~\cite{Devuyst2005}.
The segments were extracted from whole-night recordings without regard to the underlying sleep stages or the presence of artifacts.
Each segment was annotated for spindles by two human experts, except for the last two segments that were only annotated by the first expert.
Depending on the segment, the experts were presented with the C3-A1 or CZ-A1 EEG channel and were not given any information about the sleep stages identified by a different expert.
Based on the sleep stages annotated by the separate expert, we removed all spindle annotations outside N2 sleep.

The BD dataset (see also section~\ref{ssec:methods-sleep-staging}) includes spindle annotations for artifact-free N1, N2, and N3 sleep stages~\cite{Ritter2018}.
As in the original study~\cite{Ritter2018}, we focused on spindles in N2 sleep in our analyses.
The spindle annotations were created by an expert, with a second expert verifying annotations in case of uncertainty.
While the BD dataset does not specify which EEG channels were used for the annotations, it is reasonable to assume that the expert analyzed the same channels as the ones investigated in the study presenting the dataset: F3-A2, F4-A1, C3-A2, and C4-A1~\cite{Ritter2018}.

\paragraph{Model.}
Following the SUMO study, we considered the spindle detection task as a segmentation problem for single EEG channels~\cite{Kaulen2022}.
We adopted the SUMO model architecture for SUMOv2, which is a U-Net with two encoder and two decoder blocks.
SUMOv2 accepts input sequences of arbitrary length and outputs two segmentation masks that indicate for each data point in the input sequence whether a spindle is present (maximum in the first mask) or not (maximum in the second mask).
We joined consecutive indications of spindle presence to form the final spindle annotations, consisting of a starting sample and a duration.

To detect spindles using SUMOv2 or to train the model, we preprocessed the data by splitting it into contiguous blocks of N2 sleep according to sleep stage annotations.
Each block of N2 sleep was then filtered with a 20th order Butterworth high-pass filter at 0.3~Hz, followed by a 20th order Butterworth low-pass filter at 30~Hz, downsampled to 100~Hz, and normalized by subtracting the median amplitude and dividing by the interquartile range.
Amplitudes outside the range of -20 to 20 were clipped to the respective boundary.

\paragraph{Training.}
For training SUMOv2, we split the MODA dataset into a training and test set.
The test set consisted of the data of 36 subjects with 3 blocks of 115~s each (see the SUMO study for further details on how test subjects were selected~\cite{Kaulen2022}).
The training set was further split into six cross validation folds, each containing the data of five subjects with 10 blocks of 115~s each and 19 subjects with 3 blocks of 115~s each.
After hyperparameter optimization on this cross validation split, we retrained the final model on the entire training set with 10\% of the data reserved for early stopping.
The DREAMS and BD datasets were not part of the training or the validation process and were used for testing purposes only.

We trained the model using the Adam optimizer, a batch size of 12, a learning rate of 0.005, and a generalized Dice loss, which is a variant of the Dice loss that is more robust to class imbalance~\cite{Sudre2017}.
To prevent overfitting, we trained the model until the IoU-F1 score (see next section) on the validation set did not improve for 300 consecutive training epochs or until 800 training epochs were reached.

We found the original SUMO model to be sensitive to variations in EEG amplitudes, posing challenges for datasets with differing amplitude distributions, such as those from patient groups with less pronounced spindle activity or other recording setups.
To address this issue with SUMOv2, we used data augmentation by random rescaling of EEG amplitudes during training to make the model more robust.
Each sample was randomly chosen to be either upscaled (multiplied by a random factor between 1 and 2) or downscaled (multiplied by a random factor between 0.5 and 1) to ensure adaptability to varying amplitude distributions.

\paragraph{Evaluation.}
When evaluating SUMOv2 on the BD dataset, we applied the model separately to each of the four EEG channels present in the data and then aggregated the detected spindles across channels by taking the union of annotations (i.e., overlapping annotations in different channels were merged and non-overlapping annotations were kept separate).

Following standard procedures outlined in the MODA study~\cite{Lacourse2020}, we postprocessed detected spindles for all datasets by merging spindles shorter than 0.3~s and separated by less than 0.1~s, and subsequently removing spindles with a duration of less than 0.3~s or longer than 2.5~s.

To evaluate the performance of the model, we used the Intersection-over-Union (IoU) F1 score.
Given two sets of spindle annotations, the IoU-F1 score was calculated on a by-spindle basis.
Each spindle in the first set was matched with the temporally closest spindle in the second set.
If the overlap between two matched spindles divided by the duration of the combined spindles was greater than 20\%, the spindles were considered a true positive (TP).
Spindles in the first and second sets that did not have a match meeting the threshold were considered false positives (FP) and false negatives (FN), respectively.
TPs, FPs, and FNs were summed over all jointly annotated recordings or data segments (i.e., when calculating the IoU-F1 score between two experts, we summed TPs, FPs, and FNs over all recordings annotated by both experts).
The IoU-F1 score was then calculated as $2 \cdot \text{TP} / (2 \cdot \text{TP} + \text{FP} + \text{FN})$.

For the calculation of pairwise inter-rater agreement levels on the MODA dataset, we only considered expert pairs with at least five jointly annotated blocks equivalent to roughly 9.5 minutes of EEG data (280 expert pairs met this criterion).

\paragraph{Spindle Characteristics.}
For the detected spindles, we computed spindle density, duration, frequency, and amplitude.
Spindle density (spindles per minute, SPM) was determined by dividing the number of detected spindles in N2 sleep by the total duration of N2 sleep.
Spindle duration was measured as the time between the first and last sample of each spindle.
To calculate the frequency and amplitude of a spindle, we first applied a 4th-order Butterworth band-pass filter (10--16~Hz) to the unprocessed EEG signal.
Spindle frequency was calculated as the average of the instantaneous frequencies that were determined as half the reciprocals of zero-crossing intervals of the band-pass filtered signal.
Spindle amplitude was determined as the mean absolute value of the Hilbert-transformed filtered signal.

\subsection*{Ethical Approval}
The collection and analysis of the BD dataset was approved by the Institutional Review Board (IRB00001473 and IORG0001076) at the University Hospital Carl Gustav Carus, Dresden.
All other datasets were acquired from third-party databases and handled according to the relevant data sharing agreements.

\subsection*{Code Availability}
The underlying code and definitions of training/validation/test datasets for this study are available on GitHub and can be accessed via this link \url{https://github.com/dslaborg/sumov2}.
The code and model file for the RobustSleepNet model are available at \url{https://github.com/Dreem-Organization/RobustSleepNet/tree/main/pretrained_model/0dfcee73-055a-4c4d-929c-8fdf630e14f1}.

\subsection*{Data Availability}
The IS-RC dataset~\cite{Kuna2013} is available from Stephansen~et~al.~\cite{Stephansen2018} and can be accessed at \url{https://stanfordmedicine.app.box.com/s/r9e92ygq0erf7hn5re6j51aaggf50jly}.
The DREAMS dataset~\cite{Devuyst2005} is publicly available on Zenodo (\url{https://doi.org/10.5281/zenodo.2650141}).
The MODA dataset~\cite{Lacourse2020,Yetton2022} is publicly available on the Open Science Framework (\url{https://osf.io/8bma7/}).
The MASS dataset~\cite{OReilly2014}, which contains the EEG recordings, is publicly available and can be obtained from the Montreal Archive of Sleep Studies web page (\url{http://ceams-carsm.ca/mass/}).
The BD dataset is not publicly available due to patient privacy concerns but may be made available from PR (Philipp.Ritter@ukdd.de) on request.

\subsection*{Author Contributions statement}
NG and SB conceived the experiments; NG conducted the experiments; NG, SM, PR, and SB analyzed and discussed the results; NG and SB wrote the first draft of the manuscript; NG, SM, PR, and SB reviewed the manuscript.

\subsection*{Acknowledgements}
We are grateful to Justus T.C. Schwabedal for his invaluable insights and thank Martin Reißel and Volker Sander for providing us with computing resources.
This study was in part funded by the Deutsche Forschungsgemeinschaft (DFG, German Research Foundation), Project-ID 521379614 -- SFB/TRR 393 and Project-ID 454245598 -- IRTG 2773.
The funder played no role in study design, data collection, analysis and interpretation of data, or the writing of this manuscript.
Open Access funding enabled and organized by Projekt DEAL.

\subsection*{Competing interests}
The authors declare no competing interests.

\subsection*{Additional information}
Correspondence and requests for materials should be addressed to N.G. or S.B.

\end{document}